\documentclass[prb,12pt,tightenlines,showpacs,amssymb,amsmath]{revtex4}

\newtheorem{theorem}{Theorem}
\newtheorem{lemma}{Lemma}
\newtheorem{definition}{Definition}
\newtheorem{remark}{Remark}
\newtheorem{example}{Example}

\newcommand{\enlem}{\end{lemma}}
\newcommand{\closedef}{\hfill $\Diamond$ \end{definition}}
\newcommand{\enth}{\hfill $\Diamond$ \end{theorem}}
\newcommand{\encor}{\hfill $\Diamond$ \end{corollary}}
\newcommand{\enprop}{\hfill $\Diamond$ \end{proposition}}
\newcommand{\encond}{\hfill $\Diamond$ \end{condition}}
\newcommand{\exam}[1]{\begin{example}\label{ex:#1}}
\newcommand{\beremark}[1]{\begin{remark}\label{rmk:#1}}
\newcommand{\enremark}{\QED\end{remark}}

\renewcommand{\Bar}[1]{\overline{#1}}

\newcommand{\mymathbb}[1]{{\mathbb #1}} 

\newcommand{\cA}{{\cal A}}
\newcommand{\sA}{\mathsf{A}}

\newcommand{\sB}{\mathsf{B}}

\newcommand{\cC}{{\cal C}}

\newcommand{\sF}{\mathsf{F}}
\newcommand{\myF}{{\sF}}

\newcommand{\sH}{\mathsf{H}}
\newcommand{\cI}{{\cal I}}

\newcommand{\sL}{\mathsf{L}}
\newcommand{\cM}{{\cal M}}
\newcommand{\sM}{\mathsf{M}}
\newcommand{\sN}{\mathsf{N}}

\newcommand{\cQ}{{\cal Q}}

\newcommand{\cR}{{\cal R}}

\newcommand{\cT}{{\cal T}}

\newcommand{\cX}{{\cal X}}

\newcommand{\bZ}{\mymathbb{Z}}

\newcommand{\vep}{\varepsilon}
\renewcommand{\phi}{\varphi}
\renewcommand{\subset}{\subseteq}

\newcommand{\mbm}[1]{{\bf #1}}

\newcommand{\field}{\mymathbb{F}}

\newcommand{\Prob}{{\rm Pr}}

\newcommand{\tnsr}{\otimes}

\newcommand{\lag}{\langle}
\newcommand{\rag}{\rangle}
\newcommand{\crd}[1]{|#1|}
\newcommand{\bra}[1]{\lag #1 |}
\newcommand{\ket}[1]{| #1 \rag}

\newcommand{\syp}[2]{( #1,  #2 )_{\rm sp}}
\newcommand{\dmn}{d}
\newcommand{\Hch}{{\sH}}
\newcommand{\Hgn}{{\sH}}
\newcommand{\Bop}{\sL}
\newcommand{\Hcd}{\cC}
\newcommand{\Hcnd}{H_{\rm c}}

\newcommand{\Ebasis}{\sN}
\newcommand{\Ebe}{N}

\newcommand{\Icr}{J} 
\newcommand{\Fbar}{\overline{F}}
\newcommand{\Aso}{\sA}  
\newcommand{\Acn}{\sA}  
\newcommand{\Bcn}{\sB}  
\newcommand{\CPex}{\cM}
\newcommand{\CPexO}{M}
\newcommand{\Cso}{L}

\begin{document} 


\title{
A Lower Bound on the Quantum Capacity of Channels with
Correlated Errors}

\author{Mitsuru Hamada}
\email{mitsuru@ieee.org}

\affiliation{Quantum Computation and Information Project (ERATO)\\
     Japan Science and Technology Corporation,
      5-28-3, Hongo, Bunkyo-ku, Tokyo 113-0033, Japan\\
}

\date{Jan.\ 14, 2002}

\begin{abstract}
The highest fidelity
of quantum error-correcting codes of length $n$ and rate $R$
is proven to be lower bounded by
$1 - \exp [-n E(R)+ o(n)]$ for some function $E(R)$
on noisy quantum channels 
that are subject to not necessarily independent errors.
The $E(R)$ is positive below some threshold $R_0$,
which implies $R_0$ is a lower bound on the quantum capacity.
This work is an extension of the author's 
previous works [M.~Hamada, Phys.\ Rev.\ A, {\bf 65}, 052305 (2002),
e-Print quant-ph/0109114, LANL, 2001, 
and M.~Hamada, e-Print quant-ph/0112103, 
LANL, 2001], which presented the bound for channels subject to
independent errors, or channels modeled as tensor products
of copies of a completely positive linear map.
The relation of the channel class treated in this paper
to those in the previous works
are similar to that of Markov chains to
sequences of independent identically distributed random variables.
\end{abstract}

\pacs{03.67.Lx, 03.67.Hk, 89.70.+c}

\maketitle

\mbox{}
\pagebreak

\section{Introduction}

Quantum error-correcting codes 
(simply called quantum codes or {\em codes}\/ in this work) 
were discovered by Shor~\cite{shor95} and Steane~\cite{steane96} 
as schemes that protect
quantum states from decoherence during quantum computation. 
Shor~\cite{shor95} not only gave the first quantum code but also
posed a problem of determining the quantum
analog of Shannon's channel capacity.
In classical information theory, channels with independent errors 
are called memoryless channels and channels with correlated errors
are called channels with memory~\cite{Gallager}, 
which will be applied to quantum channels as well in the present work.
On quantum memoryless channels, several bounds on the quantum capacity
have been known~\cite{shor95,schumacher96, bennett96m, dss98, barnum00}, 
and also exponential convergence of fidelity of codes
was recently proved by the present author~\cite{hamada01e,hamada01g}.
It is natural to ask
whether such bounds and exponential convergence hold true or not
on channels with memory, which will be answered affirmatively in this work.

While one of the greatest incentives to investigate quantum
codes is need in quantum computing,
we are not sure which devices to use
for this purpose currently.
Hence, we do not know which channel models are appropriate,
so that treating general channels may be among what
we can proceed to now.
Thus, this paper analyzes the code performance
on a class of quantum channels that is much wider than was
treated in the literature.

In the proof of the main result below, the method of types,
which is a powerful tool from classical information theory,
plays an important role%
~\cite{csiszar_koerner,csiszar98}. 
This method was exploited 
by the Hungarian mathematician 
(information theorist) Csisz\'{a}r and coworkers 
around 1980
to present the strongest coding theorems such as the one showing the existence 
of universal channel codes asymptotically
as good as any codes~\cite{csiszar_koerner, csiszar98}.
It has 
often produced results in elementary enumerative manners,
which is also the case in this paper.

\section{Main Result for Simple Case}


As usual,
all quantum channels
and decoding (state-recovery) operations in coding systems
are described in terms of 
{\em trace-preserving completely positive}\/ (TPCP) linear maps%
~\cite{kraus71,choi75,schumacher96,barnum00,nielsen_chuang}.
Given a Hilbert space $\Hgn$ of finite dimension,
let $\Bop(\Hgn)$ denote the set of linear operators on $\Hgn$. 
In general, every CP linear map $\CPex: \Bop(\Hgn) \to \Bop(\Hgn)$ 
has an operator-sum representation 
$\CPex(\rho) = \sum_{i\in\cI} \CPexO_i \rho \CPexO_i^{\dagger}$ for some
$\CPexO_i\in\Bop(\Hgn)$, $i\in\cI$.~\cite{kraus71,choi75,schumacher96,nielsen_chuang} 
When $\CPex$ is specified by a set of operators
$\{ \CPexO_i \}_{i\in\cI}$ 
in this way, we write
$\CPex \sim \{ \CPexO_i \}_{i\in\cI}$. 

Hereafter, $\Hch$ denotes an arbitrarily fixed Hilbert space
of dimension $\dmn$, which is a prime number.
A quantum channel is a sequence of
TPCP linear maps 
$\{ \cA_n  : \Bop(\Hch^{\tnsr n}) \to \Bop(\Hch^{\tnsr n}) \}$.
We want a large subspace $\Hcd_n \subset \Hch^{\tnsr n}$
every state vector in which remains almost unchanged 
after the effect of a channel
followed by some suitable recovery operation
$ 
\cR_n: \Bop(\Hch^{\tnsr n}) \to \Bop(\Hch^{\tnsr n}). 
$ 
A pair $(\Hcd_n, \cR_n)$ consisting of such a subspace $\Hcd_n$
and a TPCP 
map $\cR_n$
is called a {\em code}\/
and its performance is evaluated in terms of minimum fidelity%
~\cite{KnillLaflamme97,dss98,barnum00}
\[
F(\Hcd_n, \cR_n\cA_n) = \min_{ \ket{\psi} \in \Hcd_n } 
\lag\psi| \cR_n\cA_n(|\psi\rag \lag\psi|) |\psi\rag,
\]
where $\cR_n\cA_n$ denotes the composition of $\cA_n$ and $\cR_n$.
Throughout, bras $\bra{\cdot}$ and kets $\ket{\cdot}$ are
assumed normalized.
A subspace $\Hcd_n$ alone is also called a code
assuming implicitly some recovery operator.

Let $F_{n,k}^{\star}(\cA_n)$ denote the
supremum of
$F(\Hcd_n, \cR_n\cA_n)$ such that there exists a code $(\Hcd_n, \cR_n)$
with $\log_{\dmn} \dim \Hcd_n \ge k$, where $n$ is a positive integer
and $k$ is a nonnegative real number.
Our goal is to estimate $F_{n,k}^{\star}(\cA_n)$ as precisely
as possible. 


First, we state the main result for an easy case,
and give a more general statement later.
Fix an orthonormal basis 
$\{ |0\rag, \dots, |\dmn-1 \rag \}$ of $\Hch$.
Put $\cX=\{0,\dots,\dmn-1 \}^2$
and $\Ebe_{(i,j)} = X^i Z^j$ for $(i,j)\in \cX$.
Here,
$X, Z \in \Bop(\Hch)$ are Weyl's unitaries,
which could be viewed as generalized Pauli operators,
and are defined by
\begin{equation}\label{eq:error_basis}
X |j \rag  = |(j-1) \bmod \dmn \, \rag, \quad
Z |j \rag = \omega^ j |j \rag,
\end{equation}
where $\omega$ is a primitive $\dmn$-th root of unity%
~\cite{weyl31,schwinger60,knill96a,knill96b,AshikhminKnill00}.
From the $\Bop(\Hch)$ basis $\{ \Ebe_{(i,j)} \}$,
we obtain a basis $\Ebasis_n = \{ \Ebe_x \mid x \in \cX^n \}$
of $\Bop(\Hch^{\tnsr n})$,
where
$ 
\Ebe_x = \Ebe_{x_1} \tnsr \dots \tnsr \Ebe_{x_n}
$ 
for $x=(x_1,\dots,x_n)\in \cX^n$.
The first channel class to be considered here
consists of those $\{ \cA_n \}$ such that
$\cA_n \sim \{ \sqrt{P_n(x)} \Ebe_{x} \}_{x\in\cX^{n}}$,
where we assume that $P_n$ are the probability distributions
of a (first-order) homogeneous Markov chain, 
i.e., that $P_n$ has the form 
\begin{equation}
 P_n(x_1,\dots, x_n)= p(x_1) \prod_{j=1}^{n-1} P(x_{j+1}| x_{j}) \label{eq:Mar}
\end{equation}
with some transition probabilities $P(v|u)$, $u,v\in\cX$,
and some initial distribution $p$.
These are generalizations of the so-called
depolarizing channel~\cite{bennett96m, dss98};
see Ruskai {\em et al.}~\cite{RuskaiSW01} for a thorough analysis of
memoryless channels with $\dmn=2$. 

Given a probability distribution $Q$ on $\cX^2$, we let
$\Bar{Q}$ and $\Bar{\Bar{Q}}$ denote the two marginal distributions:
\[
\Bar{Q}(u) = \sum_{v \in \cX}Q(u,v), \quad
\Bar{\Bar{Q}}(u) = \sum_{v \in \cX} Q(v,u), \quad u \in \cX.
\]
The classical (conditional) Kullback-Leibler information (informational divergence or relative entropy)
is denoted by $D$ and entropy by $H$.~\cite{csiszar_koerner,csiszar98,ccc} 
Specifically, for a probability distribution $Q$ on $\cX^2$,
transition (or conditional) probabilities $P(v|u)$, $u,v\in\cX$, and a probability distribution $p$ on $\cX$, 
we define $\overleftarrow{Q}(\cdot | \cdot)$ 
by $\overleftarrow{Q}(v|u)=Q(u,v)/\Bar{Q}(u)$ for $\Bar{Q}(u)>0$, 
$D(Q||P)$ 
by
\[
 D(Q||P)=\sum_{u\in\cX:\ \Bar{Q}(u)>0}\sum_{v\in\cX} Q(u,v) \log_{\dmn}
\frac{\overleftarrow{Q}(v|u)}{P(v|u)},
\]
and $H(P|p)$ by
\[
H(P|p)= - \sum_{u\in\cX:\ p(u)>0}\sum_{v\in\cX} p(u)P(v|u) \log_{\dmn}  P(v|u),
\]
which is called the entropy of $P(\cdot|\cdot)$ conditional on $p$.
We remark that $D(Q||P)$ is a conditional Kullback-Leibler information, so that
in a more consistent notation~\cite{csiszar_koerner}, it would be denoted by
$D(\overleftarrow{Q}||P|\Bar{Q})$.

By convention, we assume 
$\log(a/0) = \infty$ for $a>0$, $0\log 0= 0 \log (0/0) = 0$. 
The first form of this work's main result is the next one.
\begin{theorem}\label{th:main}
Let a channel
$\cA_n \sim \{ \sqrt{P_n(x)} \Ebe_{x} \}_{x\in\cX^n}$, $n=1,2,\dots$, 
be specified
by (\ref{eq:Mar}) with some $P(\cdot|\cdot)$ and $p$.
Then, for $0\le R \le 1$, we have
\begin{equation} \label{eq:new_exp}
\liminf_{n\to\infty} - \frac{1}{n}\log_{\dmn}[1-F^{\star}_{n,Rn}(\cA_n)]
\ge E(R,P),
\end{equation}
where 
\[
E(R,P)=\min_{Q:\, \Bar{Q}=\Bar{\Bar{Q}}} [ D(Q||P) + |1-H(\overleftarrow{Q}|\Bar{Q})-R|^+ ],
\]
$|x|^+ =\max\{x,0\}$, and
the minimization with respect to $Q$ is taken 
over all probability distributions on $\cX^2$ with 
$\Bar{Q}=\Bar{\Bar{Q}}$.
\end{theorem}
%

{\em Remarks}.\/
Roughly speaking, the theorem says
$F^{\star}_{n,Rn}(\cA_n) \gtrapprox 1- \exp_{\dmn}
[ -n E(R,P) ]$.
An immediate consequence of the theorem is that
when the Markov chain is irreducible,
the quantum capacity~\cite{shor95,schumacher96, bennett96m, dss98, barnum00} 
of $\{\cA_n \}$ is lower bounded by $1-H(P|q)$, where
$q$ is the unique stationary (steady state, or equilibrium)
distribution of the Markov chain~\cite{chung}.
To see this, observe that $E(R,P)$ is positive for $R<1-H(P|q)$
due to an easily established inequality $D(Q||P) \ge 0$ where equality occurs
if and only if $Q(u,v)=q(u)P(v|u)$ for all $u,v\in\cX$
under the constraint $\Bar{Q}=\Bar{\Bar{Q}}$.

{\em Example}.\/ 
Let us assume $\dmn=2$, 
rename the elements $(0,0), (1,0),(0,1),(1,1)$ in $\cX$ as
$\underline{0},\underline{1},\underline{2},\underline{3}$, 
and define $P(v|u)$, $u,v\in\cX$, by 
\[
P(v|u)=\begin{cases}
1- \vep & \mbox{if $u=\underline{0}$ and $v = \underline{0}$},\\
\vep/3 & \mbox{if $u = \underline{0}$ and $v \ne \underline{0}$},\\
1- \gamma & \mbox{if $u \ne \underline{0}$ and $v = \underline{0}$},\\
\gamma/3 & \mbox{if $u \ne \underline{0}$ and $v \ne \underline{0}$}.
\end{cases}
\]
In this case, $\{ A_n \}$ is analogous to the channel with memory
discussed by Gilbert~\cite{gil_chan}
in the context of classical channel coding 
(see also Gallager~\cite{Gallager}, Sec.~4.6). 
If we brought Gilbert's idea into our quantum case innocently,
we might assume $0< \vep \le \gamma <1$
and interpret $\underline{0}$ as `good state,'
$\underline{1},\underline{2},\underline{3}$ as `bad ones,'
where a state means that of the Markov chain, not a quantum state,
and $\vep$ (resp., $\gamma$) as the probability of going into a `bad state'
provided the current state be `good (resp., bad).'
For the above quantum channel, the lower bound $1- H(P|q)$ 
becomes
\[
1- \frac{
 (1-\gamma) [ h(\vep) + \vep \log_2 3 ] + 
\vep [ h(\gamma) + \gamma \log_2 3 ]
}{1-\gamma+\vep},
\]
where $h$ is the binary entropy function
$h(z)=-z \log_2 z -(1-z) \log_2 (1-z)$.
Note that when $\vep=\gamma$, the channel becomes
the depolarizing channel and the lower bound on the capacity becomes
the known one~\cite{bennett96m, hamada01e}.

\section{Proof of Theorem~1} 

\subsection{Codes Based on Symplectic Geometry}

The codes to be proven
to have the desired performance are
{\em symplectic (stabilizer, or additive) codes}\/%
~\cite{crss97,gottesman96,crss98}.
Let us recall first the basics of symplectic codes.
We can regard the index of $\Ebe_{(i,j)}=X^i Z^j$, $(i,j)\in\cX$,
as a pair of elements from the field $\myF=\field_{\dmn}=\bZ/\dmn\bZ$,
the finite field consisting of $\dmn$ elements.
Recall we put
$
\Ebe_x = \Ebe_{x_1} \tnsr \dots \tnsr \Ebe_{x_n}
$ 
for $x=(x_1,\dots,x_n)\in (\myF^2)^n$.
We write $\Ebe_{\Icr}$ for 
$\{ \Ebe_{x} \in \Ebasis_n \mid x\in\Icr \}$
where $\Icr \subset (\myF^2)^n$.
The index $((u_1,v_1),\dots,(u_n,v_n))\in (\myF^2)^n$
of a basis element can be regarded as the plain $2n$-dimensional vector
\[
x=(u_1,v_1,\dots,u_n,v_n) \in \myF^{2n}.
\]
We can equip the vector space $\myF^{2n}$ over $\myF$ with
a {\em symplectic bilinear form}\/ (symplectic paring),
which is defined by
\[
\syp{x}{y} = \sum_{i=1}^{n} u_i v_i' - v_i u_i'
\]
for the above $x$ and $y=(u'_1,v'_1,\dots,u'_n,v'_n) \in \myF^{2n}$.~\cite{artin,aschbacher,grove} 
%
Given a subspace $\Cso \subset \myF^{2n}$, let
\[
\Cso^{\perp} = \{ x \in \myF^{2n} \mid \forall y\in \Cso,\ \syp{x}{y} =0 \}.
\]

\begin{lemma} \cite{crss97,crss98} \label{lem:coset_leaders} 
Let a subspace $\Cso\subset\myF^{2n}$ satisfy 
\begin{equation*}\label{eq:self-orth}
\Cso\subset \Cso^{\perp} \quad \mbox{and} \quad \dim \Cso = n-k.
\end{equation*}
In addition, let $\Icr_0 \subset \myF^{2n}$ 
be a set satisfying
\begin{equation}\label{eq:coset_leaders}
\forall x,y\in \Icr_0,\ [ \, y-x \in \Cso^{\perp} \Rightarrow x=y \, ].
\end{equation}
Then, there exist
$\dmn^{k}$-dimensional $\Ebe_{\Icr_0}$-correcting codes.
\enlem

In fact, given a subspace $\Cso$ as above, 
there are $\dmn^{k}$ subspaces of the form
\begin{equation*}\label{eq:codespace}
\{ \psi \in \Hch^{\tnsr n} \mid \forall M\in\Ebe_{\Cso},\ M \psi =  \tau(M) \psi \},
\end{equation*}
with some scalars $\tau(M)$ (eigenvalues of $M\in \Ebe_{\Cso}$),
and each of them, together with a suitable recovery operator,
serves as an $\Ebe_{\Icr_0}$-correcting quantum code
of dimension $\dmn^k$.
Note that the direct sum of these subspaces is the whole space 
$\Hch^{\tnsr n}$.
The precise meaning of $\Ebe_{\Icr_0}$-correcting can be found,
e.g., in Knill and Laflamme~\cite{KnillLaflamme97}. 
Originally,
Lemma~\ref{lem:coset_leaders} was claimed for
the case where $\dmn=2$,
and has been generalized to the case
where $\dmn$ is a general prime%
~\cite{knill96a,knill96b,rains99,AshikhminKnill00}.

By definition, for an $\Ebe_{\Icr_0}$-correcting code $(\Hcd_n, \cR_n)$
and the channel $\{ \cA_n \}$ in the theorem,
it holds 
\begin{equation}\label{eq:fid_bound}
1-F(\Hcd) \le \sum_{x\notin {\Icr_0}} P_n(x),
\end{equation}
where $F(\Hcd)=F(\Hcd,\cR_n \cA_n)$.
We remark that, as is usually done in the literature,
it is assumed in this paper that when we speak of
an $\Ebe_{\Icr_0}$-correcting code $(\Hcd_n, \cR_n)$,
the $\cR_n$ indicates the one constructed
by Knill and Laflamme~\cite{KnillLaflamme97}.
Note that $\cR_n$ is determined from $\Icr_0$ and $\Hcd$.
The premise (\ref{eq:coset_leaders}) of Lemma~\ref{lem:coset_leaders},
is restated as that $\Icr_0$ is a set of representatives of cosets
of $\Cso^{\perp}$ in $\myF^{2n}$.
A natural choice for $\Icr_0$ would be a set
consisting of representatives each of which maximizes
the probability $P_n(x)$ in its coset%
~\cite{crss98}
since it is analogous to maximum likelihood decoding,
which is an optimum strategy for classical coding
(see Slepian~\cite{slepian56} or any textbook of information theory). 
In the proof below, 
we choose another set of representatives,
the classical counterpart of which 
(minimum entropy decoding) asymptotically yields the same performance
as maximum likelihood decoding~\cite{csiszar_koerner,CsiszarKoerner81a}.

\subsection{The Method of Types}

The theorem can be proved along the lines of
Ref.~\onlinecite{hamada01e}, which 
employed the method of types%
~\cite{csiszar_koerner,csiszar98,CsiszarKoerner81a,ccc}.
In the present case,
second-order (Markov) types rather than the usual types are used.
Needed technical tools from the method of types in the Markov case
can be found in Csisz\'{a}r {\em et al.}~\cite{ccc} and papers cited therein. 
We collect here a few basic facts on this method to be used below.

For $x=(x_1,\dots,x_n)\in\cX^n$, $n>1$,
define a probability distribution $\sM_{x}$
on $\cX^2$ by 
\[
\sM_{x}(u,v)=\frac{\crd{\{ i \mid 1\le i \le n-1, (x_i,x_{i+1}) = (u,v) \}}}{n-1}, \quad  u \in \cX,
\]
which is called the second-order {\em type}\/ or Markov type of $x$.
With $\cX$ and an element $u\in\cX$ fixed, 
the set of all possible Markov types of sequences $(x_1,\dots,x_n)$
from $\cX^n$ satisfying $x_1=u$ is denoted by $\cQ_n(\cX,u)$ 
or simply by $\cQ_n(u)$, and $\cQ_n$ stands for $\bigcup_{u\in\cX} \cQ_n(u)$.
For a type $Q\in \cQ_n(u)$, $\cT_{Q}^n(u)$ is defined as
$\{ (x_1,\dots,x_n)\in\cX^n \mid \mbox{$x_1=u$ and $\sM_{x} = Q$} \}$,
and $\cT_{Q}^n$ denotes $\bigcup_{u\in\cX} \cT_{Q}^n(u)$.

In what follows, we use
\begin{equation}\label{eq:types}
|\cT_{Q}^n(u)| \le \exp_{\dmn}[(n-1)H(\overleftarrow{Q}|\Bar{Q})], \quad u\in\cX.
\end{equation}
Note that if $x=(x_1,\dots,x_n)\in\cX^{n}$ with $x_1=u$ has type $Q$, then
$P_{n}(x)=p(u)\prod_{(a,b)\in\cX^2} P(b|a)^{(n-1)Q(a,b)}
 = p(u) \exp_{\dmn} \{ -(n-1) [H(\overleftarrow{Q}|\Bar{Q})+D(Q||P)]
\}$
and hence, (\ref{eq:types}) 
is equivalent to the latter inequality in (39) of Csisz\'{a}r {\em et al.}~\cite{ccc}, i.e.,
\begin{equation}\label{eq:types2}
\Prob \{  \sM_{\mbm{X}}=Q  \mid \mbm{X}_1=u \} 
\le \exp_{\dmn} \{ -(n-1)D(Q||P) \},
\end{equation}
where the sequence of random variables $\mbm{X}=(\mbm{X}_1, \dots, \mbm{X}_n)$
represents the Markov chain in the theorem, i.e., 
$\Prob \{ \mbm{X}_1=x_1,\dots,\mbm{X}_n=x_n \} = P_n(x_1,\dots,x_n)$
with $P_n$ defined in (\ref{eq:Mar}).
Eq.\ (\ref{eq:types}) or (\ref{eq:types2}) is a consequence of
Whittle's formula for $|\cT_{Q}^n(u)|$, a simple proof of which was given
by Billingsley~\cite{billingsley61}. The upper bound in (\ref{eq:types})
can be proved even easier
with a simple way of enumeration (Davisson {\em et al.}~\cite{dls81} or the paragraph
containing (9) of Ref.~\onlinecite{hamada99d}).

\subsection{Proof of Theorem~\protect\ref{th:main}}

The case where $R=1$ is trivial, so that we assume $R<1$ from now on.
Putting $k=\lceil Rn \rceil$,
we apply Lemma~\ref{lem:coset_leaders},
where we choose $\Icr_0$ as follows.
Assume $\dim \Cso = n-k$.
Then, $\dim \Cso^{\perp} = n+k$.\cite{artin,grove} 
For notational simplicity, we write $\Hcnd(Q)$ in place of
$H(\overleftarrow{Q}|\Bar{Q})$ for a probability distribution $Q$ on $\cX^2$.
From each of the $\dmn^{n-k}$ cosets of $\Cso^{\perp}$ in $\myF^{2n}$,
select a vector that minimizes $\Hcnd(\sM_{x})$, i.e., a vector $x$ satisfying
$\Hcnd(\sM_{x})\le \Hcnd(\sM_{y})$ for any $y$ in the coset.
This selection uses the idea of the minimum entropy 
decoder known in the classical information theory literature~\cite{CsiszarKoerner81a}.

Let $\Icr_0(\Cso)$ denote the set of the $\dmn^{n-k}$ selected vectors,
let
\[
\Aso = \{ \Cso \subset \myF^{2n} \mid \mbox{$\Cso$ linear}, \ \Cso \subset
\Cso^{\perp},\ \dim \Cso = n-k \}
\]
and for each $L\in\Aso$, 
let $\Hcd(\Cso)$ be an $\Ebe_{\Icr_0(\Cso)}$-correcting code
existence of which is ensured by 
Lemma~\ref{lem:coset_leaders}.
Putting
\[
\Fbar = \frac{1}{\crd{\Aso}} \sum_{\Cso\in\Aso} F(\Hcd(\Cso)),
\]
we will show $\liminf_n - n^{-1} \log_{\dmn}(1-\Fbar)$
$\ge E(R,P)$, which implies that, at least, one sequence of codes
has fidelity as high as promised in the theorem.
Such a method for a proof is referred to as random coding~\cite{goppa74,csiszar_koerner}.

As in the proof of Theorem~1 of Ref.~\onlinecite{hamada01e}, we have 
\begin{equation}
1- \Fbar  \le 
    \sum_{x \in \myF^{2n}} P_n(x) \frac{\crd{\Bcn(x)}}{\crd{\Aso}},
\label{eq:pr0}
\end{equation}
where
\[
\Bcn(x) = \{ \Cso \in \Aso \mid x \notin \Icr_0(\Cso) \}, \quad x\in\myF^{2n}.
\]
The fraction $\crd{\Bcn(x)}/\crd{\Aso}$ is trivially bounded as
\begin{equation} \label{eq:pr1}
\frac{\crd{\Bcn(x)}}{\crd{\Aso}} \le 1, \quad x\in\myF^{2n}.
\end{equation}
We use the next inequality%
~\cite{hamada01g}.
Let
\[
\Acn(x) = \{ \Cso \in \Aso \mid x \in \Cso^{\perp} \setminus \{ 0 \} \}.
\]
Then, $\crd{\Acn(0)}=0$ and
\begin{equation}\label{eq:C_uniform}
\frac{\crd{\Acn(x)}}{\crd{\Aso}}
 = \frac{\dmn^{n+k} - 1}{\dmn^{2n}-1}  
\le \frac{1}{\dmn^{n-k}}, \quad x\in\myF^{2n},\
 x \ne 0.
\end{equation}
This is a variant of 
the relation
established by Calderbank {\em et.\ al}\/~\cite{crss97}, or its analog
proved by Matsumoto and Uyematsu~\cite{MatsumotoUyematsu01} 
with an explicit use of the Witt lemma~\cite{artin,aschbacher} 
from the theory of bilinear forms.

Since $\Bcn(x) \subset \{ \Cso \in \Aso \mid \exists y\in\myF^{2n}, \Hcnd(\sM_y) \le \Hcnd(\sM_x), y-x\in \Cso^{\perp} \setminus \{ 0 \} \}$
from the design of $\Icr_0(\Cso)$ specified above (cf.\ Goppa~\cite{goppa74}), 
it follows that
\begin{eqnarray}
 \crd{\Bcn(x)} &\le &\sum_{y\in \myF^{2n} :\, \Hcnd(\sM_y) \le  \Hcnd(\sM_x),\ y \ne x}
 \crd{\Acn(y-x)}\notag\\
  &\le & \sum_{y\in \myF^{2n} :\, \Hcnd(\sM_y) \le  \Hcnd(\sM_x),\ y \ne x}
\crd{\Aso}{\dmn}^{-n+k}, \label{eq:pr2} 
\end{eqnarray}
where we have used (\ref{eq:C_uniform}) for the latter inequality.
Combining (\ref{eq:pr0}), (\ref{eq:pr1}) and (\ref{eq:pr2}), we 
obtain the following chain of inequalities
with the aid of the basic inequalities in (\ref{eq:types})
and (\ref{eq:types2}) as well as 
the inequality
$\min \{ a+b, 1\} \le \min \{ a, 1\} + \min \{ b, 1\}$ for $a,b \ge 0$:
{\small 
\begin{eqnarray*}
\lefteqn{1-\Fbar}\\ 
&\le& \sum_{x\in\myF^{2n}} P_n(x) \ \min \Biggl\{ \ \sum_{y\in\myF^{2n} :\, \Hcnd(\sM_y) \le  \Hcnd(\sM_x),\ y \ne x} \dmn^{-(n-k)},\ 1 \ \Biggr\}\\
  & \le & \sum_{u\in\cX} p(u) \sum_{Q\in\cQ_n(u)} \Prob \{ \sM_{\mbm{X}} = Q \mid \mbm{X}_1 = u \}
\ \min \Biggr\{ \sum_{Q'\in \cQ_n :\, \Hcnd(Q') \le \Hcnd(Q)} \frac{|\cT_{Q'}^n|}{\dmn^{n(1-R)-1}}, \ 1\ \Biggl\}\\
& \le & \dmn^3 \sum_{u\in\cX} p(u) \sum_{Q\in\cQ_n} \exp_{\dmn} [ -(n-1) D(Q || P) ] 
\sum_{Q'\in \cQ_n :\, \Hcnd(Q')\le \Hcnd(Q)} \exp_{\dmn} [ -(n-1) |1-R-\Hcnd(Q')|^{+} ]\\
& \le & \dmn^3 \sum_{Q\in\cQ_n} \exp_{\dmn} [ -(n-1) D(Q || P) ] \,
|\cQ_n| \max_{Q'\in\cQ_n :\, H(Q') \le H(Q)} \exp_{\dmn} [ -(n-1) |1-R-\Hcnd(Q')|^{+} ]\\
& \le & \dmn^3 \sum_{Q\in\cQ_n} \exp_{\dmn} [ -(n-1) D(Q || P) ] \,
|\cQ_n| \exp_{\dmn} [ -(n-1) |1-R-\Hcnd(Q)|^{+} ]\\
& \le & \dmn^3 |\cQ_n|^2 \exp_{\dmn} \{ -(n-1) \min_{Q \in \cQ_n} [ D(Q||P) 
+ |1-R-\Hcnd(Q)|^+ ] \}.
\end{eqnarray*} } 
Since $|\cQ_n|$ is polynomial in $n$,
the remaining task is to show that
\[
\liminf_{n\to\infty} \min_{Q \in \cQ_n} [ D(Q||P) 
+ |1-R-\Hcnd(Q)|^+ ]
\]
is not less than
\[
\min_{Q:\, \| \Bar{Q}-\Bar{\Bar{Q}} \| = 0 } [ D(Q||P) 
+ |1-R-\Hcnd(Q)|^+ ],
\]
which is $E(R,P)$. One sees this holds immediately
noticing that 
any $Q\in\cQ_n$ satisfies $\big\| \Bar{Q}-\Bar{\Bar{Q}} \big\| \le \frac{1}{n-1}$ 
for the norm $\| (z_1, \dots, z_{\crd{\cX}}) \| = \max_{i} |{z_i}|$,\cite{ccc}
the set of all probability distributions is compact,
and $D(Q)=D(Q||P)$ is continuous in its effective domain
$\{ Q \mid D(Q)<\infty \}$
(cf., the proof of Lemma~2 in Csisz\'{a}r {\em et al.}~\cite{ccc}).
This completes the proof.

\section{Main Result for General Case}

Theorem~\ref{th:main} actually holds for a wider class of channels.
To evaluate the fidelity of codes on a more general channel $\{ \cA_n \}$,
we first associate a sequence of probability distributions $\{ P_{\cA_n} \}$
with the channel $\{ \cA_n \}$ as in Ref.~\onlinecite{hamada01g}.
\begin{definition}\label{def:P4A}
For each $n$, let $\cA_n \sim \{ A_x^{(n)} \}_{x\in\cX^n}$,
expand $A_x^{(n)}$ as 
$A_x^{(n)} = \sum_{y\in\cX^n} a_{xy} \Ebe_y$, $x\in\cX^n$,
and define a probability distribution $P_{\cA_n}$ on $\cX^n$ by 
\[
P_{\cA_n}(y) = \sum_{x} |a_{xy}|^2, \quad y\in\cX^n.
\]
\end{definition}

{\em Example.}\/
Let $\{\cA_n\}$ be a memoryless channel
$\cA_n=\cA^{\tnsr n}$, $n=1,2,\dots$. 
It is easy to see that
$
P_{\cA_n}(y_1,\dots,y_n) = \prod_{i=1}^{n} P_{\cA}(y_i).
$

The case of memoryless channels as above
was discussed in this author's previous work~\cite{hamada01g}.
This work claims the next.
\begin{theorem}\label{th:main2}
Consider a channel
$\{ \cA_n \}$ whose
$\{ P_n=P_{\cA_n} \}$ satisfies (\ref{eq:Mar}) 
with some $P(\cdot|\cdot)$ and $p$.
Then, again, for $0\le R \le 1$, (\ref{eq:new_exp}) in Theorem~\ref{th:main}
holds.
\end{theorem}

The above theorem can be proved along the lines of this author's previous 
work~\cite{hamada01g}, which treated general memoryless quantum channels.
Namely, Theorem~\ref{th:main}
can be generalized to Theorem~\ref{th:main2} in the same
way as the result in Ref.~\onlinecite{hamada01e} was strengthened
in Ref.~\onlinecite{hamada01g}.
Here it is briefly described how to prove Theorem~\ref{th:main2}.
First, we evaluate the minimum average fidelity $F_{\rm a}(\Hcd)$,
which is another performance measure for a code $\Hcd$ 
introduced in Ref.~\onlinecite{hamada01g},
instead of the minimum fidelity $F(\Hcd)$.
Actually, we evaluate the average of $F_{\rm a}(\Hcd)$
over the whole ensemble of quantum codes
$\{ \Hcd(\Cso,i) \mid \Cso \in \Acn, 0 \le i < \dmn^{n-k} \}$,
where $\Hcd(\Cso, i)$, $i=0,\dots,\dmn^{n-k}-1$, are the $\dmn^{n-k}$
quantum codes associated with $\Cso$ as in Lemma~\ref{lem:coset_leaders};
compare the proof of Theorem~\ref{th:main} above, where
using an arbitrarily chosen code 
$\Hcd(\Cso,i)$ for each $\Cso$ was enough.
The average of $F_{\rm a}(C(L,i))$ turns out to be lower bounded
by $1- \exp_{\dmn} [ - n E(R,P) + o(n) ]$.
Then, at least, one code $\Hcd(\Cso, i)$ has this performance or higher.
As proved in Ref.~\onlinecite{hamada01g},
if we have a code with $1-F_{\rm a}(\Hcd) \le G$,
we can choose a subcode $\Hcd'$ of half the dimension with
$1-F(\Hcd') \le 2G$, which implies Theorem~\ref{th:main2}.

The major difficulty in the analysis on general channels
lay in the fact that
(\ref{eq:fid_bound}) is no longer true in the general case;
this was resolved in Ref.~\onlinecite{hamada01g} by proving
that (\ref{eq:fid_bound}) holds true 
if we replace $F(\Hcd)=F(\Hcd(L))$ by $F_{\rm a}(\Hcd(L,i))$ 
averaged over $0 \le i< d^{n-k}$.

We remark that the result of this paper readily
extends to the case where ${P_n}$ is the probability distributions of
a higher-order Markov chain. For this extension,
we have only to use higher-order types
instead of second-order types~\cite{csiszar98,ccc}.

\section{Concluding Remarks}

It should be remarked that the lower bound
$1-H(P|q)$ on the quantum capacity is not tight in general
since there is an example of a code which slightly goes beyond the bound
for some very noisy memoryless channels~\cite{dss98}.
This work, however, seems the first to demonstrate
that standard error correction 
schemes work reliably even in the presence of correlated errors
with positive information rate
for all large enough code lengths.
Moreover, the established convergence of 
the fidelity is exponential. Research in this direction is
yet to be developed in quantum information theory,
while exponent problems have already been central issues
in other fields including large-deviation theory~\cite{dembo} and
classical information theory~\cite{slepianIT,litsyn99}.

\section*{Acknowledgment}
The author would like to thank
H.~Imai and K.~Matsumoto of QCI project for support. 

\pagebreak


\begin{thebibliography}{41}
\expandafter\ifx\csname natexlab\endcsname\relax\def\natexlab#1{#1}\fi
\expandafter\ifx\csname bibnamefont\endcsname\relax
  \def\bibnamefont#1{#1}\fi
\expandafter\ifx\csname bibfnamefont\endcsname\relax
  \def\bibfnamefont#1{#1}\fi
\expandafter\ifx\csname citenamefont\endcsname\relax
  \def\citenamefont#1{#1}\fi
\expandafter\ifx\csname url\endcsname\relax
  \def\url#1{\texttt{#1}}\fi
\expandafter\ifx\csname urlprefix\endcsname\relax\def\urlprefix{URL }\fi
\providecommand{\bibinfo}[2]{#2}
\providecommand{\eprint}[2][]{\url{#2}}

\bibitem[{\citenamefont{Shor}(1995)}]{shor95}
\bibinfo{author}{\bibfnamefont{P.~W.} \bibnamefont{Shor}},
  \bibinfo{journal}{Phys.\ Rev.\ A} \textbf{\bibinfo{volume}{52}},
  \bibinfo{pages}{R2493} (\bibinfo{year}{1995}).

\bibitem[{\citenamefont{Steane}(1996)}]{steane96}
\bibinfo{author}{\bibfnamefont{A.~M.} \bibnamefont{Steane}},
  \bibinfo{journal}{Phys.\ Rev.\ Letters} \textbf{\bibinfo{volume}{77}},
  \bibinfo{pages}{793} (\bibinfo{year}{1996}).

\bibitem[{\citenamefont{Gallager}(1968)}]{Gallager}
\bibinfo{author}{\bibfnamefont{R.~G.} \bibnamefont{Gallager}},
  \emph{\bibinfo{title}{Information Theory and Reliable Communication}}
  (\bibinfo{publisher}{John Weily \& Sons}, \bibinfo{address}{NY},
  \bibinfo{year}{1968}).

\bibitem[{\citenamefont{Schumacher}(1996)}]{schumacher96}
\bibinfo{author}{\bibfnamefont{B.}~\bibnamefont{Schumacher}},
  \bibinfo{journal}{Phys.\ Rev.\ A} \textbf{\bibinfo{volume}{54}},
  \bibinfo{pages}{2614} (\bibinfo{year}{1996}), \eprint{quant-ph/9604023}.

\bibitem[{\citenamefont{Bennett et~al.}(1996)\citenamefont{Bennett, DiVincenzo,
  Smolin, and Wootters}}]{bennett96m}
\bibinfo{author}{\bibfnamefont{C.~H.} \bibnamefont{Bennett}},
  \bibinfo{author}{\bibfnamefont{D.~P.} \bibnamefont{DiVincenzo}},
  \bibinfo{author}{\bibfnamefont{J.~A.} \bibnamefont{Smolin}},
  \bibnamefont{and} \bibinfo{author}{\bibfnamefont{W.~K.}
  \bibnamefont{Wootters}}, \bibinfo{journal}{Phys.\ Rev.\ A}
  \textbf{\bibinfo{volume}{54}}, \bibinfo{pages}{3824} (\bibinfo{year}{1996}),
  \eprint{quant-ph/9604024}.

\bibitem[{\citenamefont{DiVincenzo et~al.}(1998)\citenamefont{DiVincenzo, Shor,
  and Smolin}}]{dss98}
\bibinfo{author}{\bibfnamefont{D.~P.} \bibnamefont{DiVincenzo}},
  \bibinfo{author}{\bibfnamefont{P.~W.} \bibnamefont{Shor}}, \bibnamefont{and}
  \bibinfo{author}{\bibfnamefont{J.~A.} \bibnamefont{Smolin}},
  \bibinfo{journal}{Phys.\ Rev.\ A} \textbf{\bibinfo{volume}{57}},
  \bibinfo{pages}{830} (\bibinfo{year}{1998}), \bibinfo{note}{correction:
  Phys.\ Rev.\ A, 59, p.~1717}, \eprint{quant-ph/9706061}.

\bibitem[{\citenamefont{Barnum et~al.}(2000)\citenamefont{Barnum, Knill, and
  Nielsen}}]{barnum00}
\bibinfo{author}{\bibfnamefont{H.}~\bibnamefont{Barnum}},
  \bibinfo{author}{\bibfnamefont{E.}~\bibnamefont{Knill}}, \bibnamefont{and}
  \bibinfo{author}{\bibfnamefont{M.~A.} \bibnamefont{Nielsen}},
  \bibinfo{journal}{IEEE Trans. Information Theory}
  \textbf{\bibinfo{volume}{46}}, \bibinfo{pages}{1317} (\bibinfo{year}{2000}),
  \eprint{quant-ph/9809010}.

\bibitem[{\citenamefont{Hamada}(2002)}]{hamada01e}
\bibinfo{author}{\bibfnamefont{M.}~\bibnamefont{Hamada}},
  \bibinfo{journal}{Phys.\ Rev.\ A} \textbf{\bibinfo{volume}{65}},
  \bibinfo{pages}{052305} (\bibinfo{year}{2002}); \bibinfo{note}{e-Print,
  quant-ph/0109114, LANL, 2001}, \eprint{quant-ph/0109114}.

\bibitem[{\citenamefont{Hamada}(2001)}]{hamada01g}
\bibinfo{author}{\bibfnamefont{M.}~\bibnamefont{Hamada}},
  \bibinfo{type}{e-Print} \bibinfo{number}{quant-ph/0112103},
  \bibinfo{institution}{LANL} (\bibinfo{year}{2001}). \bibinfo{note}{Submitted
  to IEEE Trans. Information Theory}.

\bibitem[{\citenamefont{Csisz\'{a}r and
  K\"{o}rner}(1981{\natexlab{a}})}]{csiszar_koerner}
\bibinfo{author}{\bibfnamefont{I.}~\bibnamefont{Csisz\'{a}r}} \bibnamefont{and}
  \bibinfo{author}{\bibfnamefont{J.}~\bibnamefont{K\"{o}rner}},
  \emph{\bibinfo{title}{Information Theory: Coding Theorems for Discrete
  Memoryless Systems}} (\bibinfo{publisher}{Academic}, \bibinfo{address}{NY},
  \bibinfo{year}{1981}{\natexlab{a}}).

\bibitem[{\citenamefont{Csisz\'{a}r}(1998)}]{csiszar98}
\bibinfo{author}{\bibfnamefont{I.}~\bibnamefont{Csisz\'{a}r}},
  \bibinfo{journal}{IEEE Trans. Information Theory}
  \textbf{\bibinfo{volume}{IT-44}}, \bibinfo{pages}{2505}
  (\bibinfo{year}{1998}).

\bibitem[{\citenamefont{Kraus}(1971)}]{kraus71}
\bibinfo{author}{\bibfnamefont{K.}~\bibnamefont{Kraus}},
  \bibinfo{journal}{Annals of Physics} \textbf{\bibinfo{volume}{64}},
  \bibinfo{pages}{311} (\bibinfo{year}{1971}).

\bibitem[{\citenamefont{Choi}(1975)}]{choi75}
\bibinfo{author}{\bibfnamefont{M.-D.} \bibnamefont{Choi}},
  \bibinfo{journal}{Linear Algebra and Its Applications}
  \textbf{\bibinfo{volume}{10}}, \bibinfo{pages}{285} (\bibinfo{year}{1975}).

\bibitem[{\citenamefont{Nielsen and Chuang}(2000)}]{nielsen_chuang}
\bibinfo{author}{\bibfnamefont{M.~A.} \bibnamefont{Nielsen}} \bibnamefont{and}
  \bibinfo{author}{\bibfnamefont{I.~L.} \bibnamefont{Chuang}},
  \emph{\bibinfo{title}{Quantum Computation and Quantum Information}}
  (\bibinfo{publisher}{Cambridge University Press},
  \bibinfo{address}{Cambridge, UK}, \bibinfo{year}{2000}).

\bibitem[{\citenamefont{Knill and Laflamme}(1997)}]{KnillLaflamme97}
\bibinfo{author}{\bibfnamefont{E.}~\bibnamefont{Knill}} \bibnamefont{and}
  \bibinfo{author}{\bibfnamefont{R.}~\bibnamefont{Laflamme}},
  \bibinfo{journal}{Phys.\ Rev.\ A} \textbf{\bibinfo{volume}{55}},
  \bibinfo{pages}{900} (\bibinfo{year}{1997}), \eprint{quant-ph/9604034}.

\bibitem[{\citenamefont{Weyl}(1950)}]{weyl31}
\bibinfo{author}{\bibfnamefont{H.}~\bibnamefont{Weyl}},
  \emph{\bibinfo{title}{The Theory of Groups and Quantum Mechanics}}
  (\bibinfo{publisher}{Dover}, \bibinfo{address}{NY}, \bibinfo{year}{1950}),
  \bibinfo{note}{translation from the second German ed., 1931}.

\bibitem[{\citenamefont{Schwinger}(1960)}]{schwinger60}
\bibinfo{author}{\bibfnamefont{J.}~\bibnamefont{Schwinger}},
  \bibinfo{journal}{Proc.\ Nat.\ Acad.\ Sci.\ USA}
  \textbf{\bibinfo{volume}{46}}, \bibinfo{pages}{570} (\bibinfo{year}{1960}).

\bibitem[{\citenamefont{Knill}(1996{\natexlab{a}})}]{knill96a}
\bibinfo{author}{\bibfnamefont{E.}~\bibnamefont{Knill}},
  \bibinfo{type}{e-Print} \bibinfo{number}{quant-ph/9608048},
  \bibinfo{institution}{LANL} (\bibinfo{year}{1996}{\natexlab{a}}).

\bibitem[{\citenamefont{Knill}(1996{\natexlab{b}})}]{knill96b}
\bibinfo{author}{\bibfnamefont{E.}~\bibnamefont{Knill}},
  \bibinfo{type}{e-Print} \bibinfo{number}{quant-ph/9608049},
  \bibinfo{institution}{LANL} (\bibinfo{year}{1996}{\natexlab{b}}).

\bibitem[{\citenamefont{Ashikhmin and Knill}(2001)}]{AshikhminKnill00}
\bibinfo{author}{\bibfnamefont{A.}~\bibnamefont{Ashikhmin}} \bibnamefont{and}
  \bibinfo{author}{\bibfnamefont{E.}~\bibnamefont{Knill}},
  \bibinfo{journal}{IEEE Trans. Information Theory}
  \textbf{\bibinfo{volume}{47}}, \bibinfo{pages}{3065} (\bibinfo{year}{2001}),
  \eprint{quant-ph/0005008}.

\bibitem[{\citenamefont{Ruskai et~al.}(2001)\citenamefont{Ruskai, Szarek, and
  Werner}}]{RuskaiSW01}
\bibinfo{author}{\bibfnamefont{M.~B.} \bibnamefont{Ruskai}},
  \bibinfo{author}{\bibfnamefont{S.}~\bibnamefont{Szarek}}, \bibnamefont{and}
  \bibinfo{author}{\bibfnamefont{E.}~\bibnamefont{Werner}},
  \bibinfo{type}{e-Print} \bibinfo{number}{quant-ph/0101003},
  \bibinfo{institution}{LANL} (\bibinfo{year}{2001}).

\bibitem[{\citenamefont{Csisz\'{a}r et~al.}(1987)\citenamefont{Csisz\'{a}r,
  Cover, and Choi}}]{ccc}
\bibinfo{author}{\bibfnamefont{I.}~\bibnamefont{Csisz\'{a}r}},
  \bibinfo{author}{\bibfnamefont{T.~M.} \bibnamefont{Cover}}, \bibnamefont{and}
  \bibinfo{author}{\bibfnamefont{B.-S.} \bibnamefont{Choi}},
  \bibinfo{journal}{IEEE Trans. Information Theory}
  \textbf{\bibinfo{volume}{IT-33}}, \bibinfo{pages}{788}
  (\bibinfo{year}{1987}).

\bibitem[{\citenamefont{Chung}(1967)}]{chung}
\bibinfo{author}{\bibfnamefont{K.~L.} \bibnamefont{Chung}},
  \emph{\bibinfo{title}{Markov Chains With Stationary Transition
  Probabilities}} (\bibinfo{publisher}{Springer}, \bibinfo{address}{NY},
  \bibinfo{year}{1967}), \bibinfo{edition}{2nd} ed.

\bibitem[{\citenamefont{Gilbert}(1960)}]{gil_chan}
\bibinfo{author}{\bibfnamefont{E.~N.} \bibnamefont{Gilbert}},
  \bibinfo{journal}{The Bell System Technical Journal}
  \textbf{\bibinfo{volume}{39}}, \bibinfo{pages}{1253} (\bibinfo{year}{1960}).

\bibitem[{\citenamefont{Calderbank et~al.}(1997)\citenamefont{Calderbank,
  Rains, Shor, and Sloane}}]{crss97}
\bibinfo{author}{\bibfnamefont{A.~R.} \bibnamefont{Calderbank}},
  \bibinfo{author}{\bibfnamefont{E.~M.} \bibnamefont{Rains}},
  \bibinfo{author}{\bibfnamefont{P.~W.} \bibnamefont{Shor}}, \bibnamefont{and}
  \bibinfo{author}{\bibfnamefont{N.~J.~A.} \bibnamefont{Sloane}},
  \bibinfo{journal}{Phys.\ Rev.\ Lett.} \textbf{\bibinfo{volume}{78}},
  \bibinfo{pages}{405} (\bibinfo{year}{1997}), \eprint{quant-ph/9605005}.

\bibitem[{\citenamefont{Gottesman}(1996)}]{gottesman96}
\bibinfo{author}{\bibfnamefont{D.}~\bibnamefont{Gottesman}},
  \bibinfo{journal}{Phys.\ Rev.\ A} \textbf{\bibinfo{volume}{54}},
  \bibinfo{pages}{1862} (\bibinfo{year}{1996}), \eprint{quant-ph/9604038}.

\bibitem[{\citenamefont{Calderbank et~al.}(1998)\citenamefont{Calderbank,
  Rains, Shor, and Sloane}}]{crss98}
\bibinfo{author}{\bibfnamefont{A.~R.} \bibnamefont{Calderbank}},
  \bibinfo{author}{\bibfnamefont{E.~M.} \bibnamefont{Rains}},
  \bibinfo{author}{\bibfnamefont{P.~W.} \bibnamefont{Shor}}, \bibnamefont{and}
  \bibinfo{author}{\bibfnamefont{N.~J.~A.} \bibnamefont{Sloane}},
  \bibinfo{journal}{IEEE Trans.\ Inform.\ Theory}
  \textbf{\bibinfo{volume}{44}}, \bibinfo{pages}{1369} (\bibinfo{year}{1998}),
  \eprint{quant-ph/9608006}.

\bibitem[{\citenamefont{Artin}(1957)}]{artin}
\bibinfo{author}{\bibfnamefont{E.}~\bibnamefont{Artin}},
  \emph{\bibinfo{title}{Geometric Algebra}} (\bibinfo{publisher}{Interscience
  Publisher}, \bibinfo{address}{New York}, \bibinfo{year}{1957}).

\bibitem[{\citenamefont{Aschbacher}(2000)}]{aschbacher}
\bibinfo{author}{\bibfnamefont{M.}~\bibnamefont{Aschbacher}},
  \emph{\bibinfo{title}{Finite Group Theory}} (\bibinfo{publisher}{Cambridge
  University Press}, \bibinfo{address}{Cambridge, UK}, \bibinfo{year}{2000}),
  \bibinfo{edition}{2nd} ed.

\bibitem[{\citenamefont{Grove}(2001)}]{grove}
\bibinfo{author}{\bibfnamefont{L.~C.} \bibnamefont{Grove}},
  \emph{\bibinfo{title}{Classical Groups and Geometric Algebra}}
  (\bibinfo{publisher}{American Mathematical Society},
  \bibinfo{address}{Providence, Rhode Island}, \bibinfo{year}{2001}).

\bibitem[{\citenamefont{Rains}(1999)}]{rains99}
\bibinfo{author}{\bibfnamefont{E.~M.} \bibnamefont{Rains}},
  \bibinfo{journal}{IEEE Trans.\ Information Theory}
  \textbf{\bibinfo{volume}{45}}, \bibinfo{pages}{1827} (\bibinfo{year}{1999}),
  \eprint{quant-ph/9703048}.

\bibitem[{\citenamefont{Slepian}(1956)}]{slepian56}
\bibinfo{author}{\bibfnamefont{D.}~\bibnamefont{Slepian}},
  \bibinfo{journal}{The Bell System Technical Journal}
  \textbf{\bibinfo{volume}{35}}, \bibinfo{pages}{203} (\bibinfo{year}{1956}),
  \bibinfo{note}{reprinted in E.\ R.\ Berlekamp, ed., {\em Key Papers in The
  Development of Coding Theory},\/ NY, IEEE Press, 1974.}

\bibitem[{\citenamefont{Csisz\'{a}r and
  K\"{o}rner}(1981{\natexlab{b}})}]{CsiszarKoerner81a}
\bibinfo{author}{\bibfnamefont{I.}~\bibnamefont{Csisz\'{a}r}} \bibnamefont{and}
  \bibinfo{author}{\bibfnamefont{J.}~\bibnamefont{K\"{o}rner}},
  \bibinfo{journal}{IEEE Trans. Information Theory}
  \textbf{\bibinfo{volume}{IT-27}}, \bibinfo{pages}{5}
  (\bibinfo{year}{1981}{\natexlab{b}}).

\bibitem[{\citenamefont{Billingsley}(1961)}]{billingsley61}
\bibinfo{author}{\bibfnamefont{P.}~\bibnamefont{Billingsley}},
  \bibinfo{journal}{Ann. Math. Statist.} \textbf{\bibinfo{volume}{32}},
  \bibinfo{pages}{12} (\bibinfo{year}{1961}).

\bibitem[{\citenamefont{Davisson et~al.}(1981)\citenamefont{Davisson, Longo,
  and Sgarro}}]{dls81}
\bibinfo{author}{\bibfnamefont{L.~D.} \bibnamefont{Davisson}},
  \bibinfo{author}{\bibfnamefont{G.}~\bibnamefont{Longo}}, \bibnamefont{and}
  \bibinfo{author}{\bibfnamefont{A.}~\bibnamefont{Sgarro}},
  \bibinfo{journal}{IEEE Trans. Information Theory}
  \textbf{\bibinfo{volume}{IT-27}}, \bibinfo{pages}{431}
  (\bibinfo{year}{1981}).

\bibitem[{\citenamefont{Han and Hamada}(1999)}]{hamada99d}
\bibinfo{author}{\bibfnamefont{T.~S.} \bibnamefont{Han}} \bibnamefont{and}
  \bibinfo{author}{\bibfnamefont{M.}~\bibnamefont{Hamada}},
  \bibinfo{journal}{IEEE Trans. Information Theory}
  \textbf{\bibinfo{volume}{IT-45}}, \bibinfo{pages}{756}
  (\bibinfo{year}{1999}).

\bibitem[{\citenamefont{Goppa}(1974)}]{goppa74}
\bibinfo{author}{\bibfnamefont{V.~D.} \bibnamefont{Goppa}},
  \bibinfo{journal}{Problems of Information Transmission}
  \textbf{\bibinfo{volume}{10}}, \bibinfo{pages}{89} (\bibinfo{year}{1974}).

\bibitem[{\citenamefont{Matsumoto and Uyematsu}(2001)}]{MatsumotoUyematsu01}
\bibinfo{author}{\bibfnamefont{R.}~\bibnamefont{Matsumoto}} \bibnamefont{and}
  \bibinfo{author}{\bibfnamefont{T.}~\bibnamefont{Uyematsu}},
  \bibinfo{type}{e-Print} \bibinfo{number}{quant-ph/0105151},
  \bibinfo{institution}{LANL} (\bibinfo{year}{2001}).

\bibitem[{\citenamefont{Demobo and Zeitouni}(1998)}]{dembo}
\bibinfo{author}{\bibfnamefont{A.}~\bibnamefont{Demobo}} \bibnamefont{and}
  \bibinfo{author}{\bibfnamefont{O.}~\bibnamefont{Zeitouni}},
  \emph{\bibinfo{title}{Large Deviations Techniques and Applications}}
  (\bibinfo{publisher}{Springer}, \bibinfo{address}{Berlin},
  \bibinfo{year}{1998}), \bibinfo{edition}{2nd} ed.

\bibitem[{\citenamefont{Slepain}(1973)}]{slepianIT}
\bibinfo{editor}{\bibfnamefont{D.}~\bibnamefont{Slepain}}, ed.,
  \emph{\bibinfo{title}{Key Papers in The Development of Information Theory}}
  (\bibinfo{publisher}{IEEE Press}, \bibinfo{address}{NY},
  \bibinfo{year}{1973}).

\bibitem[{\citenamefont{Litsyn}(1999)}]{litsyn99}
\bibinfo{author}{\bibfnamefont{S.}~\bibnamefont{Litsyn}},
  \bibinfo{journal}{IEEE Trans. Information Theory}
  \textbf{\bibinfo{volume}{IT-45}}, \bibinfo{pages}{385}
  (\bibinfo{year}{1999}).

\end{thebibliography}
\end{document}